# Two-chip power-scalable THz-generating semiconductor disk laser

HEYANG GUOYU,[1,2] CHRISTIAN KRISO,[1] FAN ZHANG,[1] MATTHIAS WICHMANN,[1] WOLFGANG STOLZ,[1] KSENIA A. FEDOROVA,[1] AND ARASH RAHIMI-IMAN[1,*]

[1]*Department of Physics and Materials Sciences Center, Philipps-Universität Marburg, Marburg, 35032, Germany*
[2]*College of Applied Sciences, Beijing University of Technology, Beijing, 100124, China*
*\*Corresponding author: a.r-i@physik.uni-marburg.de*

**We demonstrate a compact two-chip terahertz-emitting vertical-external-cavity surface-emitting laser (TECSEL) source, which provides 1-THz output based on intracavity frequency conversion of dual-wavelength emission in a periodically-poled lithium niobate crystal. The type-I frequency conversion scheme at room temperature highly benefits from the power-scaling possibilities in a multi-chip cavity with intracavity powers in excess of 500 W.**

Optically pumped semiconductor disk lasers (SDLs), also known as vertical-external-cavity surface-emitting lasers (VECSELs) have been demonstrated within the last decade with enormous high-power multimode and single-frequency continuous-wave (cw) output from a single semiconductor chip [1-3]. SDLs are also capable of high-performance mode-locked operation [4-8] and intra-cavity frequency conversion with a wide coverage of wavelengths based on available materials [9,10], for which semiconductor technologies have been mastered in the past decades [11]. All recent demonstrations render VECSELs capable of being configured for a plethora of operation modes and being employed for various applications [12-16].

Remarkably, intracavity frequency conversion has been even reported for VECSELs to deliver terahertz output from a nonlinear crystal – commonly using periodically-poled lithium niobate (PPLN) – at room temperature based on two-color operation [17], exploiting the achievable high-intracavity intensities. Indeed, the performance and limitations of the so-called THz-generating VECSEL (TECSEL) have been investigated and discussed in the literature (see [12] and references therein). Consecutively and gradually, the development of alternative cavity designs and device configurations took place [18-20].

To further improve the performance and wavelength flexibility of intracavity frequency conversion with VECSELs, two-chip devices have been developed. While the wavelength spacing in the two-color operation mode can be conveniently set by an intracavity etalon of proper thickness, the gain spectra in two-chip systems can be tuned against each other naturally by the design of the chips' active region in order to broaden the accessible dual-wavelength range. In contrast, for one-chip TECSEL, the separation of the two lasing wavelengths is limited by the gain bandwidth of the single VECSEL chip and can hardly be increased beyond a few nanometers, unless SDLs with chirped quantum-dot layers in the active region, which provide inhomogeneously-broadened gain spectra, are employed [21]. In this context, a T-cavity configuration, which exploits two VECSEL chips with different emission wavelengths, has been developed by Hessenius et al. [22]. Such platform was demonstrated to deliver 300 W intracavity power for frequency conversion schemes [19]. Yet, due to the fact that the two laser modes are orthogonally polarized, this configuration is only suitable for type-II frequency conversion, which has its own drawbacks in terms of performance. However, type-I frequency conversion schemes, for which both laser modes are co-linearly polarized, render the serially-connected two-chip VECSELs attractive candidates for dual-wavelength tunability and power scaling purposes [20,23]. In such configuration, gain spectra of the involved chips can be conveniently tuned against each other for even the same chip design by means of the on-chip angle, owing to the incident-angle-dependent detuning effect in planar microcavities [24].

In this work, we demonstrate a power-scalable THz laser at room temperature based on intracavity difference-frequency generation (DFG) in an SDL system, utilizing a flexible, compact, serially-connected two-chip cavity design. In contrast to a single-chip THz-generating SDL, we can double the THz output from the same PPLN crystal irradiated by the high-power dual-wavelengths cavity light, which exhibits a single polarization and a desired wavelength separation. Furthermore, we present a long-time-trace measurement of the stable, room-temperature cw-THz output.

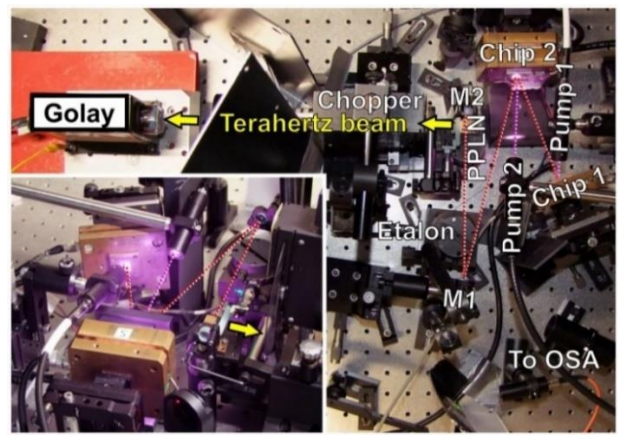

Fig. 1. Image of the two-chip laser setup with components labeled. Dotted lines indicate the 1010-nm intracavity beam and the 808-nm pump beams in red and pink, respectively. Optical paths are also indicated in the inset, which shows a close-up view of the Z-shaped cavity. The THz signal is directed through a chopper wheel towards a Golay cell using two polymer lenses, approximately at the locations from which yellow arrows originate.

Two MOVPE-grown VECSEL chips are employed in our experiments. Both chips consist of 10 InGaAs quantum wells (QWs), equally spaced by GaAsP barrier layers. The resonant periodic gain (RPG) structure is arranged by overlapping the QWs with the antinodes of the standing light field. 22.5 AlAs/GaAs layer pairs form the distributed Bragg reflector (DBR). Both chips have the same design and therefore laser operation occurs at around 1010 nm. Each chip is flip-chip-bonded to a diamond heat spreader via solid-liquid interdiffusion bonding and then mounted to a water-cooled copper heat sink.

Both chips are individually pumped by fiber-coupled 808-nm diode lasers. In order to ensure operation in the $TEM_{00}$ mode, the size of the pump spot on each chip is chosen to be 15% smaller than the size of the calculated fundamental cavity mode of 400 µm and 500 µm on chip 1 and 2, respectively, for the given resonator configuration. The laser resonator is completed by serially connecting the first chip (1) – acting as a high-reflectivity (HR) plane mirror, the second chip (2) with a folding angle of 40°, an HR concave mirror (M1, radius of curvature 250 mm, 0.05% transmittance), and an HR plane mirror (M2). The total length of the Z-shaped cavity is about 52 cm. The PPLN crystal is placed close to the HR mirror (M2), where the Gaussian-beam waist (diameter) of the laser mode becomes as narrow as about 210 µm. Figure 1 displays photographs of the experimental TECSEL setup.

A 100-µm quartz etalon is inserted in the cavity at its Brewster's angle. When an etalon with suitable free spectral range (FSR) at given angle is used, the filtering effect enables two laser mode packages to oscillate simultaneously in the cavity. The above concept based on two-chip VECSELs has been previously demonstrated to deliver access to a wide range of wavelength spacings and showed efficient frequency conversion (sum-frequency and second-harmonic generation, SFG and SHG, respectively) in a PPLN crystal [20].

In order to collect the THz signal obtained via DFG in the PPLN crystal (poling angle of 67.4°, the crystal is 10 mm long, 5 mm wide and 1 mm thick), a cylindrical polyethylene lens (f = 5 mm) is used to collimate the (vertically) divergent THz beam and then a spherical lens (f = 60 mm) is used to focus it onto a Golay cell's detector aperture. It should be mentioned that we used the same etalon, PPLN crystal, curved and plane HR mirrors for, both, a single-chip and two-chip TECSEL. In the single-chip configuration, a cavity length of 50 cm with the chip 1 as an end mirror in a V-shaped cavity was used, with an arm length of 20 cm (30 cm) for the HR-mirror (chip) side. The folding angle on the curved HR mirror amounted to 8°. The diameter of the pump spot was set to be 300 µm, which is 10% smaller than the cavity mode, to ensure the $TEM_{00}$-mode operation for efficient DFG in the PPLN crystal. In this cavity, the crystal was similarly placed close to the plane HR mirror, where the Gaussian-beam waist (diameter) reaches its minimum of 220 µm. Here, the design was adopted from previous experiments with such a PPLN crystal [25,26].

For comparison, power curves have been recorded for both chips individually and also combined in the two-chip system. In order to study the output performance of both gain chips, two VECSELs were consecutively set up in an equivalent V-shaped cavity configuration, in which the gain chip (1 or 2) acts as the folding mirror (pump size of 400 µm), with approximately equal folding angles (42° and 45°, respectively). Each single-chip VECSEL consisted of the same HR mirror (M2, 6-cm arm length), the respective gain chip and the same (HR) curved mirror (radius of curvature 250 mm, M1, 15-cm arm length).

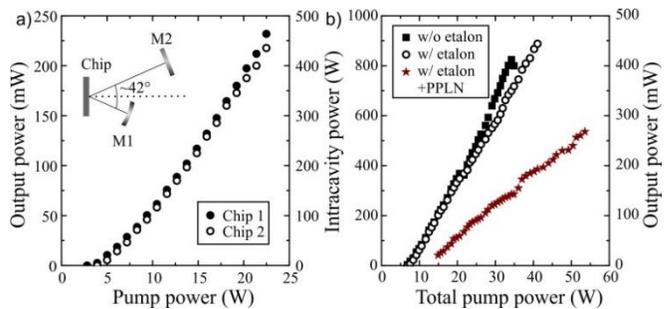

Fig. 2: a) Lasing curves for chip 1 and 2 in V-shaped cavities with HR mirrors showing output power (left) and intracavity power (right axis) as a function of the pump power. b) Two-chip power curves are displayed for an empty cavity (i.e. without "w/o" etalon, squares), with "w/" etalon (circles) as well as with etalon and PPLN crystal (stars).

As shown in Fig. 2a, the output performance of both VECSELs is comparable, with intracavity powers up to about 450 W in the free-running regime (i.e. without etalon). As two chips are combined in the aforementioned Z-shaped cavity configuration and both pumped, the gain of the two similar chips add up to a total gain much stronger than that of the single-chip devices. This leads to intracavity powers in excess of 820 W (free running), even when the etalon is inserted (up to 900 W, in the single-color mode), as shown in Fig. 2b.

Since the FSR of the etalon matches the separation needed for 1 THz output (~3.5 nm at 1010 nm), one can achieve efficient frequency conversion in the dual-wavelength regime. To deduce intracavity powers in any lasing configuration for given pump powers, we monitor the total output behind the curved HR mirror (M1, 0.05% transmittance) of our single/two-chip TECSEL. The net pump powers are generally ~30% lower (according to the reflected pump light) in these VECSELs than the gross pump powers.

The corresponding dual-wavelength spectrum for THz generation was recorded for every measured power setting. The spacing between the two laser lines (colors) was actively kept for each pump setting to be around 3.5 nm by adjusting the etalon, achieving 3.50 ± 0.03 nm (~ 1.025 ± 0.010 THz in this wavelength region) throughout the measurements described in the following. Representative spectra recorded by an *ANDO AQ-6315A* optical spectrum analyzer (OSA) with 0.1-nm resolution are shown in Fig. 3a, at intracavity powers of 150, 250 and 400 W. It is worth noting that the THz output power changed only little when the two modes had significantly different peak intensities (up to 50 %) compared to the case with the both modes having approximately the same intensity.

Figure 3b shows the corresponding SHG and SFG signals together with the fundamental laser lines which were recorded when the 10-Hz chopped THz signal was in parallel detected by the Golay cell and monitored as a signal trace by an oscilloscope (*Tektronix TDS 1002*). The green curve in the composite spectrum (with the intensity scale independent from that of the fundamental light) shows simultaneously-generated short-wavelength light with two SHG peaks surrounding the central SFG peak. The latter one, i.e. sum-frequency signal, clearly shows that even if the intensities of the two laser lines do not have the same amplitude, the intensity of SFG is still stronger than that of SHG. This indicates that both colors lase simultaneously and stable frequency mixing (SFG and DFG) is obtained; otherwise, only SHG would be visible as the sole nonlinear frequency-conversion effect in the PPLN crystal.

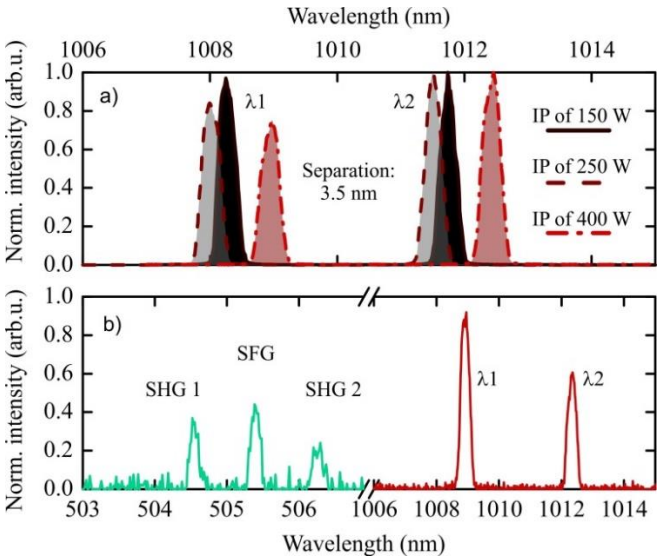

Fig. 3: a) Optical spectrum of two-color laser operation for three different intracavity powers. The wavelength spacing between the mode packages in the two-chip TECSEL is maintained at 3.50 ± 0.03 nm (~ 1.025 ± 0.010 THz in this wavelength region). b) A two-color spectrum at the fundamental wavelength together with the corresponding SHG and SFG signals at 250 W of intracavity power.

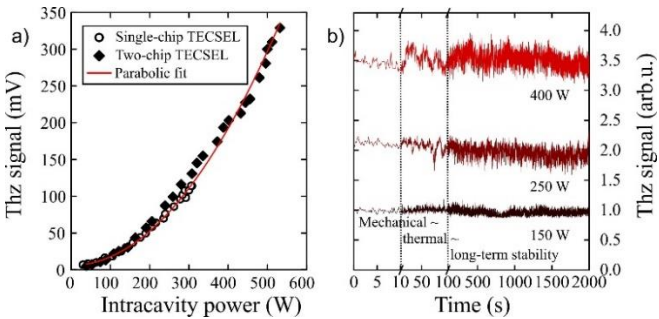

Fig. 4: a) Input-output curve of the two-chip TECSEL (closed rhombs) compared to that of a comparable single-chip device (open circles) using the same components. The red line corresponds to a parabolic fit with a slight vertical offset to account for the low-intensity noise level. The symbol sizes are comparable to the measurement uncertainty in our detection configuration. b) THz-signal stability measured over a time span of 2000 seconds. For clarity, different time windows are displayed, indicating the influence of mechanical, thermal and long-term effects on the device's output. The signal in each case is normalized to the signal level at 150 W intracavity power.

To characterize our two-chip TECSEL, we have performed both input-output measurements and long-time stability measurements. Here, the laser-line spacing of 3.50 ± 0.03 nm was maintained during all measurements.

Firstly, a number of performance curves (THz signal vs. intracavity power) were recorded, one of which is exemplarily shown in Fig. 4a (filled symbols). Remarkably, our device exhibits a high dual-wavelength stability, repeatability and robustness. The signal output compares well with previous work using a similar crystal [26]. However, the signal could not be translated to absolute power owing to the lack of a gauged Golay cell. Nevertheless, based on the estimated intracavity circulating power of about 540 W, and according to the results demonstrated in Ref. [17], it can be expected that our system can generate THz output in excess of 650 µW. A parabolic fit indicates a quadratic intensity dependence expected for the nonlinear frequency conversion in the PPLN crystal. Most striking is the fact, that by combining two chips, from each of which TECSEL operation with certain equivalent output levels can be individually obtained (cf. open-symbol curve), the generated THz signal can be indeed (technically) more than doubled for significantly less-than-twice the intracavity-power level. For both cases, if we increased the pump power further, SHG became much stronger and THz was no longer generated. This can be attributed to jumps from the $TEM_{00}$ to a multi-mode operation at higher pump powers.

Secondly, the THz signal was continuously recorded by a high-performance oscilloscope (*LeCroy WavePro 760Zi-A*) for 2000 seconds to study the signal stability of THz output via DFG over different time scales. THz-signal time traces recorded for three different intracavity powers, 150 W, 250 W and 400 W, respectively, are shown in Fig. 4b. Here, the time traces (obtained under 10-Hz chopping rate) are shown on three different time scales. The first time scale (0-10 s) accounts for mechanical noise and exhibits almost no fluctuations, i.e. the THz output is stable, as well as constant, and DFG is equally efficient. For the second time scale (10-100 s), some fluctuations can be observed, particularly at higher pump rates) and it is possibly caused by thermal fluctuations caused by the air conditioner and chiller. On the long time scale, it is worth noting that the TECSEL worked continuously for more than four hours to complete the whole stability measurement.

Furthermore, we experienced that the two-chip configuration requires far less adjustment of the etalon for the maintenance of two-color operation during power-measurement series. Only, the etalon is readjusted from time to time in such series in order to keep the wavelength spacing as close as possible to the target value. This renders the two-chip configuration a powerful platform for intracavity frequency conversion in general and DFG for THz generation in particular.

In summary, we demonstrated an efficient, stable and improved cw-THz-generating room-temperature laser system by serially connecting two similar VECSEL chips in one cavity. Our overall THz-signal level is drastically increased in comparison to the single-chip TECSEL due to the doubled gain and therefore much higher intracavity powers in contrast to an equivalent single-chip configuration with its natural performance limitations. In addition, stability measurements reveal reliable, robust and efficient operation at various intracavity-power levels. This renders two-chip TECSELs promising room-temperature sources for high-power cw-THz radiation, which can be readily obtained by semiconductor-disk-laser technology in combination with PPLN crystals and employed for applications such as THz imaging and spectroscopy [27-32].

**Funding.** Deutsche Forschungsgemeinschaft (DFG) (RA2841/1-1), China Scholarship Council, European Union's Horizon 2020 research and innovation programme under the Marie Sklodowska-Curie actions (EU H2020 MSCA) (789670).

**Acknowledgment**. The authors would like to thank NAsP III/V GmbH for the fabrication and bonding of the laser chips.

# References

1. B. Heinen, T.-L. Wang, M. Sparenberg, A. Weber, B. Kunert, J. Hader, S. W. Koch, J. V. Moloney, M. Koch, and W. Stolz, Electron. Lett **48**, 516 (2012).
2. T.-L. Wang, B. Heinen, J. Hader, C. Dineen, M. Sparenberg, A. Weber, B. Kunert, S.W. Koch, J. V. Moloney, M. Koch, and W. Stolz, Laser Photonics Rev. **6**, L12 (2012).
3. F. Zhang, B. Heinen, M. Wichmann, C. Möller, B. Kunert, A. Rahimi-Iman, W. Stolz, and M. Koch, Opt. Express **22**, 12817 (2014).
4. M. A. Gaafar, A. Rahimi-Iman, K. A. Fedorova, W. Stolz, E. U. Rafailov, and M. Koch, Adv. Opt. Photonics **8**, 370 (2016).
5. B. W. Tilma, M. Mangold, C. A. Zaugg, S. M. Link, D. Waldburger, A. Klenner, A. S. Mayer, E. Gini, M. Golling, and U. Keller, Light-Sci Appl. **4**, e310 (2015).
6. M. Scheller, T.-L. Wang, B. Kunert, W. Stolz, S. W. Koch, and J. V. Moloney, Electron. Lett **48**, 588 (2012).
7. D. Waldburger, S. M. Link, M. Mangold, C. G. E. Alfieri, E. Gini, M. Golling, B. W. Tilma, and U. Keller, Optica, **3**, 844 (2016).
8. A. Laurain, I. Kilen, J. Hader, A. R. Perez, P. Ludewig, W. Stolz, S. Addamane, G. Balakrishnan, S. W. Koch, and J. V. Moloney, Appl. Phys. Lett., **113**, 121113 (2018).
9. J. Lin, H. M. Pask, D. J. Spence, C. J. Hamilton, and G. P. A. Malcolm, Opt. Express **20**, 5219 (2012).
10. R. Bek, S. Baumgärtner, F. Sauter, H. Kahle, T. Schwarzbäck, M. Jetter, and P. Michler, Opt. Express **23**, 19947 (2015).
11. M. Guina, A. Rantamäki, and A. Härkönen, J. Phys. D. Appl. Phys. **50**, 383001 (2017).
12. A. Rahimi-Iman, J. Opt. **18**, 093003 (2016).
13. S. M. Link, A. Klenner, M. Mangold, C. A. Zaugg, M. Golling, B. W.Tilma, and U. Keller, Opt. Express **23**, 5521 (2015).
14. M. Link, D. J. H. C. Maas, D. Waldburger, and U. Keller, Science, **356**, 1164 (2017).
15. K. G.Wilcox, F. Rutz, R. Wilk, H. D. Foreman, J. S. Roberts, J. Sigmund, H. L. Hartnagel, M. Koch, and A. C. Tropper, Electron. Lett., **42**, 1159 (2006).
16. E. Kantola, A. Rantamaki, I. Leino, J. P. Penttinen, T. Karppinen, S. R. Mordon, and M. Guina, IEEE. J. Sel. Top. Quant., **25**, (2019).
17. M. Scheller, J. M. Yarborough, J. V. Moloney, M. Fallahi, M. Koch, and S. W. Koch, Opt. Express **18**, 27112 (2010).
18. J. R. Paul, M. Scheller, A. Laurain, A. Young, S. W. Koch, and J. Moloney, Opt. Lett. **38**, 3654 (2013).
19. M. Lukowski, C. Hessenius, and M. Fallahi, IEEE J. Sel. Top. Quantum Electron. **21**, 1 (2015).
20. F. Zhang, M. Gaafar, C. Möller, W. Stolz, M. Koch, and A. Rahimi-Iman, IEEE Photonics Technol. Lett. **28**, 927 (2016).
21. E.U. Rafailov, M. A. Cataluna, and W. Sibbett, Nat. Photonics **1**, 395 (2007).
22. C. Hessenius, M. Lukowski, and M. Fallahi, Appl. Phys. Lett. **101**, 121110 (2012).
23. L. Fan, M. Fallahi, J. Hader, A. R. Zakharian, J. V. Moloney, W. Stolz, S. W. Koch, R. Bedford, and J. T. Murray, Appl. Phys. Lett. **90**, 181124 (2007).
24. F. Zhang, C. Möller, M. Koch, S. W. Koch, A. Rahimi-Iman, and W. Stolz, Appl. Phys. B. **123**, 291 (2017).
25. M. Wichmann, M. Stein, A. Rahimi-Iman, S. W. Koch, and M. Koch, J. Infrared Milli. Terahz. Waves **35**, 503 (2014).
26. F. Zhang, S. M. Wang, A. Rehn, M. Wichmann, G. Urbasch, A. Rahimi-Iman, S. W. Koch, and M. Koch, J. Infrared Milli. Terahz. Waves **37**, 536 (2016).
27. X. Yin, B. W.-H. Ng, and D. Abbott, New York: Springer, 2012
28. J.-H. Son, CRC Press, 2014.
29. M. Koch, Springer Series in Optical Sciences, **85**. Springer, Berlin, Heidelberg, 2003.
30. K. J. Siebert, T. Löffler, H. Quast, M. Thomson, T. Bauer, R. Leonhardt, S. Czasch, and H. G. Roskos, Phys. Med. Biol. **47**, 3743 (2002).
31. A. J. Fitzgerald, E. Berry, N. N. Zinovev, G. C. Walker, M. A. Smith, and J. M. Chamberlain, Physics in Medicine and Biology, **47**, 67 (2002).
32. P. Knobloch, C. Schildknecht, T. Kleine-Ostmann, M. Koch, S. Hoffmann, M. Hofmann, E. Rehberg, M. Sperling, K. Donhuijsen, G. Hein, and K. Pierz, Phys. Med. Biol. **47**, 3857 (2002).